\documentclass[12pt]{article}
\usepackage{epsfig, amsmath}

\def\beq{\begin{equation}}
\def\eeq#1{\label{#1}\end{equation}}
\def\eeqn{\end{equation}}

\def\beqa{\begin{eqnarray}}
\def\eeqa#1{\label{#1}\end{eqnarray}}
\def\eeqan{\end{eqnarray}}

\let\bar=\overbar

\def\O{{\cal O}}

\def\Dslash{\not{\hbox{\kern-4pt $D$}}}
\def\dslash{\not{\hbox{\kern-2pt $\del$}}}

\def\msb{{\bar{\ssstyle M \kern -1pt S}}}

\def\Title#1{\begin{center} {\Large {\bf #1} } \end{center}}

\begin{document}

\Title{Collisionless Shocks -- Magnetic Field Generation and Particle Acceleration}

\bigskip\bigskip

%+\addtocontents{toc}{{\it D. Reggiano}}
%+\label{ReggianoStart}

\begin{raggedright}  

{\it J. Trier Frederiksen, C. B. Hededal, T. Haugb\o lle, \AA . Nordlund, \index{JTF_CH_TH_AN}\\
Niels Bohr Institute, Dept. of Astronomy\\
Juliane Maries Vej 30, 2100 K\o benhavn \O , DENMARK}
\bigskip\bigskip
\end{raggedright}

\begin{abstract}

We present numerical results from plasma particle simulations of 
collisionless shocks and ultra-relativistic counter-streaming plasmas.
We demonstrate how the field-particle interactions lead to 
particle acceleration behind the shock-front. The post-shock particle 
energy spectra are found to be segmented power laws. Specifically, we 
observe power law indices consistent with $p \approx -2.2$ for high gamma particles, 
and $p \approx 0.5$ for low gamma particles. The break is found at a gamma higher 
by $\sim 0.5 M_i/m_e$ than the gamma of the up-stream plasma relative to the shock.
Further, we find that ultra relativistic counter-streaming plasmas create 
large scale magnetic fields and that the generated field propagate at 
$v\sim{}c$. The magnetic field generation is due to a Weibel-like two-stream 
instability.  These results may help explain the origin of the magnetic fields
and accelerated electrons responsible for afterglow synchrotron radiation from gamma ray bursts.

%Still, we stress that conditions for application of Weibel's theory are 
%violated in highly relativistic collisionless shocks, and that,
%consequently, scalings of the field generation mechanism and instabilities
%could be different than in the standard Weibel case.
%%AA  The abstract is supposed to be a summary, so should only contain things
%%AA  that are actually mentioned in the paper.

\end{abstract}

\section{Introduction} 

One requirement in validating MHD models of a plasma is that collisions 
are able to maintain the distribution function close to a Maxwellian. 
However, in the context of the  external shock model of gamma ray 
burst after-glows \cite{bib:meszaros_rees} the mean free path for collisions 
is much larger than the depth of the fireball shell. Thus, collisions 
are so rare that the distribution functions departs radically from 
Maxwellians. In the reference frame of the shock 
one observes two counter-streaming plasmas. In the absence of 
collisions this gives rise to a deep interpenetration of two 
streaming near-delta-function populations. Such a situation clearly
requires a kinetic treatment of the plasma.

The existence of a strong magnetic field in the expanding fireball shell 
is required in order to explain the observed synchrotron radiation. 
Medvedev \& Loeb
\cite{bib:MedvedevLoeb} showed through a linear kinetic treatment how a
two-stream magnetic instability (a generalization of the Weibel
instability \cite{bib:Weibel,bib:YoonDavidson}) can generate a
strong magnetic field ($\epsilon_B \sim 10^{-5}$-$10^{-1}$ of
equipartition energy density) in the collisionless shock front
(see also discussion in \cite{bib:RossiRees}). We
note in passing that this instability is well-known in other plasma
physics disciplines (e.g. laser-plasma interactions \cite{bib:califano1}
\cite{bib:YangGallantAronsLangdon}).

A power-law distributed population of relativistic electrons is also 
required to produce the observed synchrotron radiation spectra. 
All together, we therefore identify the following questions as being
essential in the discussion of collisionless shocks and GRB afterglows:
1) Which mechanism is responsible for the particle 
acceleration in collisionless plasma shocks -- and what implications can 
be inferred from the answer to this question?
2) Can one possibly avoid {\it ab initio} 
assumptions of a magnetic field in GRB after-glows, and instead come
up with a picture where the magnetic field is selfconsistently generated,
using known and defensible physics?

\section{Simulation Tool} 

Simulations were performed using a self-consistent time-averaged implicit 
fully 3D electromagnetic particle-in-cell code compliant with special 
relativity. The PIC-code was originally written by Dr. Michael Hesse 
for simulating reconnection topologies \cite{bib:HesseKuzenova}, and has
been redeveloped and enhanced by the present authors to obey special relativity 
and to be fully ${\mathcal O}$($\Delta t^2$) + ${\mathcal O}$($\Delta 
x^2$).

The code solves the Maxwell equations for the electromagnetic 
field $$ \partial_t {\bf E} = c^2 \left( \partial_{\bf x} \times {\bf{B}} 
- \mu_0 \bf{J} \right) ,\hspace{25pt} \partial_t {\bf B} = - \partial_{\bf 
x} \times {\bf E} ,\hspace{25pt}  \partial_{\bf x} \cdot {\bf B} = 0 
,\hspace{25pt} \partial_{\bf x} \cdot {\bf E} =  \rho_c / \epsilon_0 .$$

\noindent Particle velocities and positions are defined continuously 
throughout phase space. Fields and field source terms are defined on a 
fully 3D Yee lattice \cite{bib:Yee}. The sources in Maxwell's equations  
are formed by weighted averaging of particle data to the field grid,
using quadratic spline weighted interpolation (Triangular Shaped Cloud
scheme).

The particle-to-mesh (source) and mesh-to-particle (force) interpolations have 
identical weighting schemes, in order to obey momentum conservation and 
eliminate particle self-forces through coupling to the grid 
\cite{bib:HockneyEastwood}.  The ion to electron mass ratio is typically 
chosen in the range $m_{0i}/m_{0e} \sim 8-25$.
The particle equations of motion are integrated by solving, for all
particles, $$\partial_t(\gamma(v_i) 
{\bf v}_i) = q_i m^{-1}_{0i} \left( {\bf E} + {\bf v}_i \times {\bf B} 
\right) ,\hspace{40pt} \partial_t{\bf r}_i = {\bf v}_i.$$ 

A typical grid size is $\{n_x,n_y,n_z\} = \{50, 50, 400\}$, with about
25 particles/cell or $\sim 10^7$ particles. 
Typical dimensions $L_\perp \sim 4\delta_i \sim 16\delta_e$ and 
$L_z \sim 25 \delta_i \sim 100\delta_e$, chosen so as to resolve a 
significant number of ion skin-depths $\delta_e$ -- vital to our 
conclusions.  
\newpage
\section{Magnetic Field Generation}

A small anisotropic perturbation to an isotropic probability density
distribution for a plasma gives rise to a plasma instability that generates
a magnetic field \cite{bib:Weibel} (with a relativistic generalization by
\cite{bib:YoonDavidson}). A similar two-stream situation is described in
detail by Medvedev and Loeb \cite{bib:MedvedevLoeb}; two counter-streaming
plasmas penetrate each other with a relative velocity and relativistic bulk
gamma factor $\Gamma_{sh}$.  In the presence of an infinitesimal magnetic
seed field, electrons and ions deflect into separate channels.  This creates
currents that amplify the existing magnetic configuration, producing an
instability and magnetic field amplification.

From the requirement that the total plasma momentum should be conserved, and since $$ 
{\bf P}_{tot} = \sum_i {\bf p}_i({\bf r}_i)_{_V} + \int_{V} {\bf S}({\bf 
r}) ~d{\bf r}, $$ the (electro)magnetic field produced by the two-stream 
may acquire part of the momentum lost by the entire two-stream population 
in the shock; this opens the possibility that magnetic field structures 
created in the shock migrate downstream of the shock and thus 
carry away some of the momentum impinging on the shock.
Here, ${\bf{S}}$ denotes the Poynting flux in a volume $V$, and 
${\bf p}_i$ are the particle momenta in $V$.

Our experiments show that this does indeed happen; the magnetic field 
structures created in the shock are also carried along downstream.   
Since the impinging beam of particles looses energy only gradually 
downstream of their first encounter with the denser plasma, there is
ample energy present for  sustaining and amplifying the field further.

We find that, as the instability develops downstream, the perpendicular
size of the structures increase systematically with distance from the 
shock.  The growth mechanism is somewhat similar to the merging of
smaller cells into larger ones below the surface of a strongly stratified
convection zone in a star.  Smaller cells are advected by larger cells,
and their constituent structures merge into those of the larger cells.
\begin{figure}[!h] 
\begin{center}
\epsfig{figure=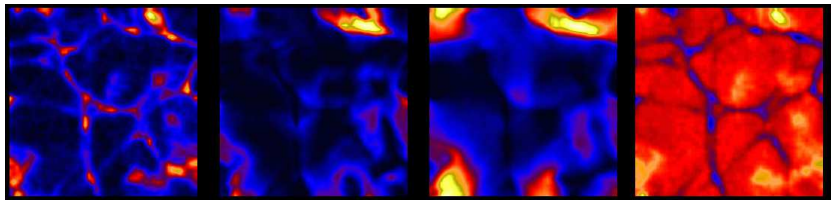,width=0.72\textwidth} 
\caption{Slices at constant $z \sim 0.75 z_{max}$ showing: $\rho_i$,
 $|B_{\perp}|$,$|B_{\perp}|+|E_{\perp}|$, and $J_z$.}
\label{fig:Slice}
\end{center} 
\end{figure}

Figure \ref{fig:Slice} shows the resulting patterns, in a slice at 75\% of
the covered downstream region.  Note that, even though the patterns started
out on scales similar to the skin depths, they grow to cover essentially
the whole box width in our model.
 
\section{Particle Acceleration}

If the correlation length of the electro-magnetic field is smaller than  
typical particle gyro radii the field is able to work as an efficient 
scatter-mechanism for the particles. Coupling of the two streams through 
fluctuating {\bf B}- 
and {\bf E}-fields then provide a mechanism for heating the quiescent 
population (in either reference frame).
Also, it will act as a channel to transfer and equalize 
the kinetic energy between ions and electrons in each of the two beams. 
Hence, electrons will be accelerated strongly 
($\gamma_e\sim{}\frac{1}{2}\gamma_{sh}m_{0i}/m_{0e})$. Of course, some of the 
kinetic energy must go to the production of electromagnetic turbulence, to 
provide the scatter mechanism in the first place. 

There is thus the intriguing possibility that a chain of energy transfers
between various energy pools may provide a selfconsistent picture where
both magnetic field generation and particle acceleration is explained:

Initially the energy resides predominantly in the bulk kinetic energy of
the heavier protons (and ions).  The fluctuating magnetic field created by 
the two-stream instability helps scatter some of the incoming protons, 
and also carries away some of the bulk momentum.  The scattered protons
create a fluctuating electric field, which tends to equilibrate the energy
between protons and electrons, thus accelerating the electrons.  This 
mechanism is qualitatively different from Fermi acceleration in that
acceleration is provided {\em in situ} in the down-stream plasma, rather 
than by scattering of particles back and forth across the shock.

\begin{figure}[ht!]
\begin{center}
\epsfig{file=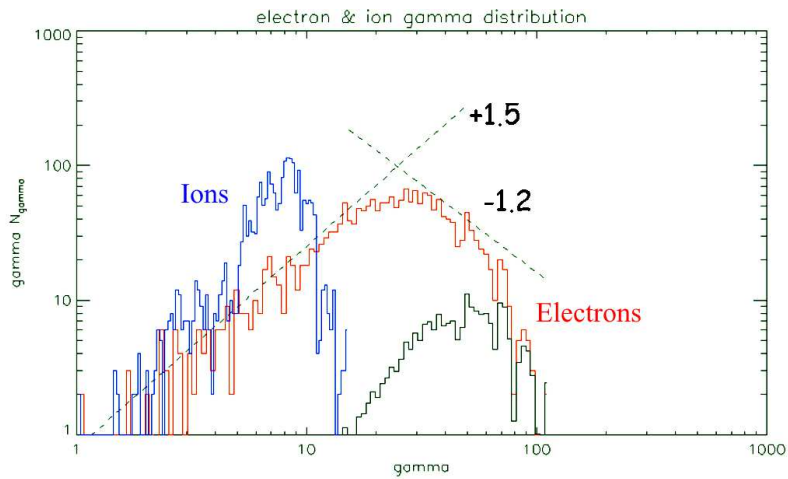,width=0.7\textwidth}
\caption{Probability Distribution Functions as produced by the collisionless 
shock in one of our experiments.}
\label{fig:Sims}
\end{center}
\end{figure}

The Probability Distribution Functions illustrated in Fig.\ \ref{fig:Sims}
shows that this chain  of events actually takes place, and that the mechanism
is capable of producing a population of accelerated electrons.

\section{Conclusions}
A central point to be made here is that magnetic field 
generation and particle acceleration may well
be generic and unavoidable in collisionless relativistic shocks.
If so, the ratios $\epsilon_B$ and $\epsilon_e$ are both results 
of the same process, and cannot be regarded as independent,
free parameters.

Qualitatively, the electro-magnetic field acts as the `catalyst'
that allows the desired `reaction' to proceed; the bulk kinetic
energy in the up-stream plasma needs to be converted to randomized
particle kinetic energy while maintaining the bulk impulse, for
the conventional shock picture to be applicable.  A priori, it
is unclear how this happens in a collisionless shock.  
The scenario proposed here, inspired by the results of the simulations, 
offers a likely 
mechanism for the thermalization of the bulk kinetic energy, which 
automatically also provides a natural explanation for the
presence of a strong magnetic field and a power law population 
of accelerated electrons.

One may hope that further studies along these lines will provide
quantitative predictions for the fractions of the bulk kinetic 
energy that go into magnetic field energy and energy of accelerated 
electrons, $\epsilon_B$ and $\epsilon_e$, respectively; these
are the parameters than are needed / assumed in conventional GRB 
afterglow shock modeling.
However,
what is ultimately observed is the resulting synchrotron-like 
radiation spectrum.  Thus, rather than first trying to abstract
scalar parameters $\epsilon_B$ and $\epsilon_e$ from the simulations
(where these ratios may be expected to actually depend on the distance 
from the shock) it may be a better approach to compute synthetic
radiation spectra directly from the models, and to use scaling
laws to predict what would be observed from the corresponding,
real afterglow shocks.

\bigskip
We are grateful to Dr. Michael Hesse / GSFC for generously providing the 
original PIC code and for helpful discussion on the implementation and 
numerical issues in particle simulations.

\def\Discussion{
\setlength{\parskip}{0.3cm}\setlength{\parindent}{0.0cm}
     \bigskip\bigskip      {\Large {\bf Discussion}} \bigskip}
\def\speaker#1{{\bf #1:}\ }
\def\endDiscussion{}

\vspace*{-12pt}
\Discussion

\speaker{Berger}
Is it possible to explain the wide range of p values which is inferred to 
be about 1.4 - 2.8 from afterglow observation?

\speaker{Nordlund} 
That remains to be seen; we have so far carried out only a few experiments.

\speaker{Brandenburg} 
Very near the shock front, the ion and electron densities were very nearly
stagnant  in the lateral direction; only further away did the structures 
wobble.What is it that holds electron and ion concentrations so nearly 
fixed near the
shock?

\speaker{Nordlund} 
The feature itself.  This IS the first encounter of the beam with any
resistamce, which creates a standing perturbation.

\speaker{Rossi}
Do your result indicate that we should change the theory that we use to 
fit afterglow? If yes, how?

\speaker{Nordlund}
We should certainly not assume constant $\epsilon_B$ and $\epsilon_e$ 
to hold in the entire shocked plasma, but rather develop an understanding
of their evolution and mutual dependence.

\endDiscussion

\end{document}